# Corrosion Resistance of Sulfur-Selenium Alloy Coatings

Sandhya Susarla[1]*, Govind Chilkoor[2]*, Yufei Cui[3]*, Taib Arif[4], Thierry Tsafack[1], Anand Puthirath[1], Parambath M. Sudeep[4], Jawahar R. Kalimuthu[2], Aly Hassan[4], Samuel Castro-Pardo[5], Morgan Barnes[1], Rafael Verduzco[1,6], Tobin Filleter[4], Nikhil Koratkar[7,8], Venkataramana Gadhamshetty[2,9‡], Muhammad M Rahman[1‡], Pulickel M. Ajayan[1‡].

[1]Department of Materials Science and NanoEngineering, Rice University, TX 77005, USA.

[2]Department of Civil and Environmental Engineering, South Dakota School of Mines and Technology, SD 57701, USA.

[3]Department of Bioengineering, Rice University, TX 77005, USA.

[4]Department of Mechanical and Industrial Engineering, University of Toronto, Ontario M5S 3G8, Canada.

[5]Department of Chemistry, Rice University, TX 77005, USA.

[6]Department of Chemical and Biomolecular Engineering, Rice University, TX 77005, USA.

[7]Department of Mechanical, Aerospace and Nuclear Engineering, Rensselaer Polytechnic Institute, NY 12180, USA.

[8]Department of Materials Science and Engineering, Rensselaer Polytechnic Institute, NY 12180, USA.

[9]2D materials for Biofilm Engineering Science and Technology (2D-BEST), South Dakota School of Mines and Technology, SD 57701, USA

‡Correspondence to: Venkata.Gadhamshetty@sdsmt.edu, mr64@rice.edu, ajayan@rice.edu
* Equal contribution

**Abstract**

Despite decades of research, metallic corrosion remains a long-standing challenge in many engineering applications. Specifically, designing a material that can resist corrosion both in abiotic as well as biotic environments remains elusive. Here we design a lightweight sulfur-selenium (S-Se) alloy with high stiffness and ductility that can serve as a universal corrosion-resistant coating with protection efficiency of ~99.9% for steel in a wide range of diverse environments. S-Se coated mild steel shows a corrosion rate that is 6-7 orders of magnitude lower than bare metal in abiotic (simulated seawater and sodium sulfate solution) and biotic (sulfate-reducing bacterial medium) environments. The coating is strongly adhesive and mechanically robust. We attribute the high corrosion resistance of the alloy in diverse environments to its semi-crystalline, non-porous, anti-microbial, and viscoelastic nature with superior mechanical performance, enabling it to successfully block a variety of diffusing species.

**One Sentence Summary:** A lightweight and mechanically robust 'Sulfur-Selenium' alloy as an anti-corrosive universal coating.

Corrosion has compromised structural materials since the dawn of human civilization. The annual direct cost of metallic corrosion is estimated to be ~$300 billion in the United States and ~€200 billion in Europe (*1*). Steel is one of the most used metals in production and manufacturing, utilities, transportation, defense, and infrastructure; however, it is susceptible to a diverse range of corrosive environments catalyzed by oxygen, moisture, electrolytes, and microbes (*2*, *3*). It remains a long-standing challenge to develop a universal coating that can protect steel in both abiotic as well as biotic environments. For example, while inorganic coatings (e.g., zinc- and chromium-based) form an effective barrier against moisture and $Cl^-$ ions, they do not work well against sulfate-reducing biofilms (*1*, *4*). Moreover, the utilization of such inorganic coatings is being increasingly restricted by governmental regulation because of their adverse effects on human health and the environment. Polymer-based coatings such as polyaniline, polyurethane, zeolite, and epoxy protect steel effectively under abiotic conditions (*5–9*); however, they are susceptible to hydrolysis in biotic environments that induce cracks and localized corrosion (*10*). This biotic corrosion arises from the organic components of polymer coatings which serve as preferential macronutrients for cell metabolism, thereby supporting microbial growth and subsequent microbial-induced corrosion. Anti-microbial properties can be imparted to polymer coatings by dispersing nanoparticles ($TiO_2$, Ag) in the base matrix (*11*, *12*). While such antimicrobial coatings are beneficial for combating corrosion under abiotic/biotic conditions, the addition of nanoparticles based on noble metals greatly increases the cost of the polymer coating.

Another emerging strategy to develop combined abiotic and biotic corrosion resistive films is to use 2D coatings of layered graphene or hexagonal boron nitride (*13–16*). Conformal coatings based on 2D materials and their heterostructures offer attractive routes to protect metals against aggressive environments including thermal oxidation, microbial corrosion, and atmospheric corrosion. However, the current nanomanufacturing methods, high production costs, and limited availability make it challenging to achieve defect (pinhole) free growth of 2D material coatings on large scale surfaces in a cost-effective manner (*17*, *18*). Therefore, there is an urgent need for an effective, universal, and inexpensive corrosion-resistant coating to protect metals such as steel from a diverse array of corrosive species, both under biotic and abiotic conditions.

Here we report an inexpensive anti-corrosive coating using a stable alloy comprised of two chalcogenides- sulfur and selenium (S-Se) that are cast as a conformal ductile coating on mild steel substrates. This S-Se alloy adheres well to steel, acts as a corrosion-resistant coating under both abiotic (simulated seawater and 0.1 M sodium sulfate salt media) and biotic (an aggressive sulfate-reducing bacterial medium) environments, is non-porous and has superb ductility and stiffness combinations when compared to conventional anti-corrosive coatings. Further, these chalcogens possess inherent anti-microbial properties against both Gram-negative and Gram-positive bacterial strains (*19*, *20*) and do not dissolve in typical solvents including water, acetone and isopropanol, making it an ideal anti-corrosive coating for many disparate environments. Ab initio analyses of the interaction of the alloy with both the surface of steel and corrosion agents, confirm its excellent adhesion to the substrate and its superior resistance to various corrosion agents. As such, our results demonstrate that the as-produced S-Se coating shows outstanding potential as a high-performance and versatile barrier coating in a variety of corrosive environments.

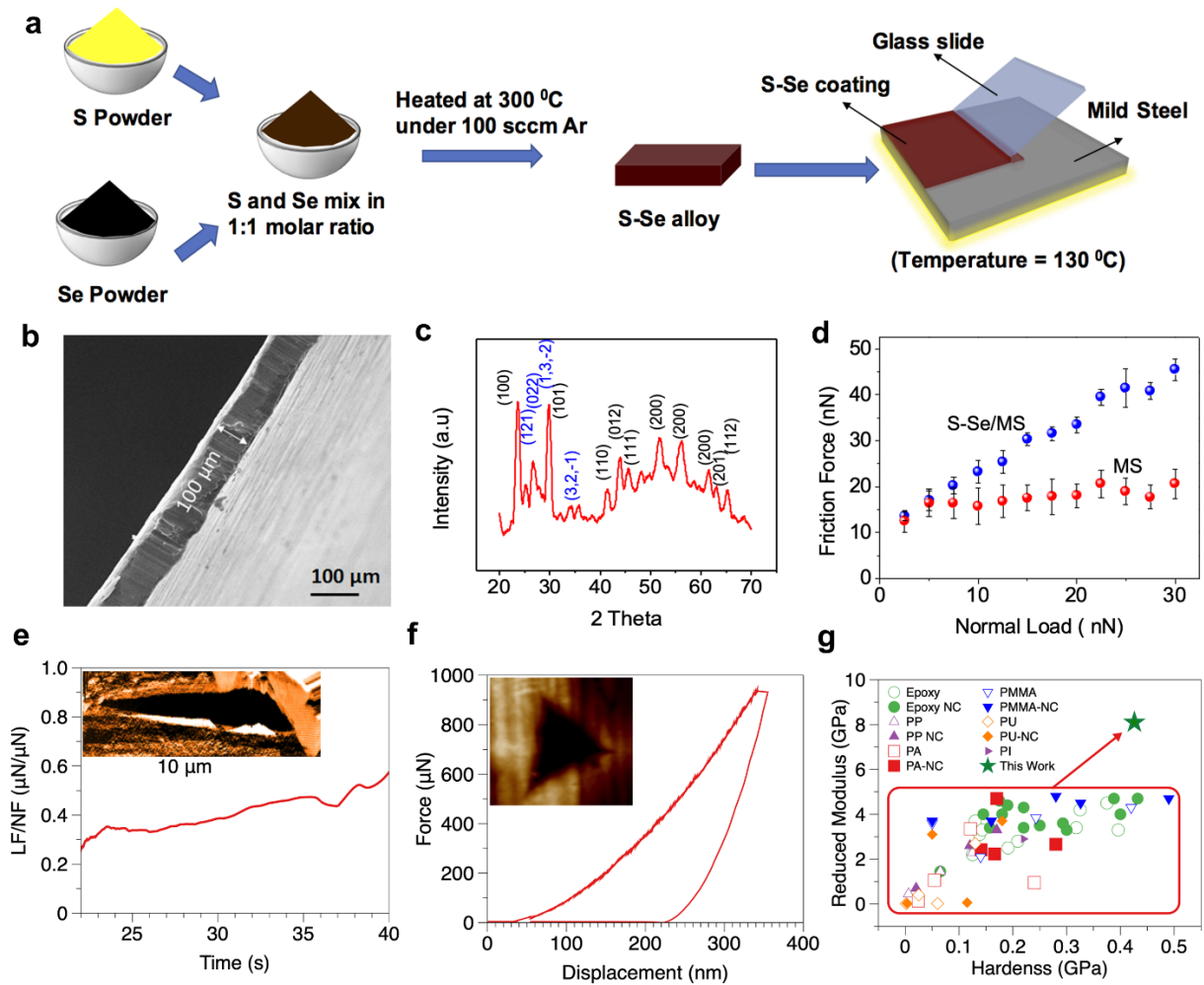

**Figure 1: Synthesis, structure, tribological, and mechanical properties of S-Se coating**.

(a) Schematic of the coating procedure of S-Se on Mild steel (MS). (b) Cross-section SEM of S-Se/MS sample indicating the thickness of the alloy coating to be ~100 $\mu$m. (c) XRD of S-Se coating indicating its semi-crystalline nature. Peaks indexed in blue ink correspond to monoclinic S-Se and those in black ink correspond to hexagonal Se. (d) Friction force on S-Se/MS and MS samples at relative humidity (RH) of ~18% using a sharp silicon tip. (e) Friction coefficient of S-Se coatings as a function of time from nano-scratch testing (inset shows the progressive scratch in the sample). (f) Load-displacement curve of S-Se coating from nanoindentation testing (inset shows a typical single indent in the sample). (g) Comparison of the S-Se alloy coating with state-of-the-art polymer and polymer nanocomposite (NC) coatings reported from nanoindentation (*21*, *22*)

**Fabrication, Structure, and Mechanical Property Profile**

Sulfur-selenium (S-Se) alloy coatings were synthesized by mixing elemental sulfur and selenium in a 1:1 molar ratio and then heating to ~300 °C in a ceramic mold placed inside a tube furnace in an inert argon atmosphere. Interestingly, the alloy behaves like an elastomeric polymer where it undergoes high deformation upon compressive loading and recovers its original shape upon removal of load at room temperature (*23*). However, it is not possible to form a uniformly thin coating onto a steel substrate at room temperature due to the temperature-dependent viscosity of this material (**Figure S1,** Supplementary Information). As the temperature of S-Se is increased from ~100 °C to ~180 °C, the viscosity is reduced by 3 orders of magnitude, and we identified the most suitable temperature (~130 °C) for coating based on the viscosity or flowability of the alloy. Thin coatings of the alloy were obtained by using a handmade doctor blade method where S-Se was kept on a hotplate at ~130 °C and coated onto mild steel (MS, AISIS 1018) coupon (**Figure 1a**). The coated MS was immediately taken out and placed at room temperature to prevent the agglomeration of the S-Se alloy owing to its high viscosity. The average thickness of the coating, as observed from the cross-sectional SEM images was determined to be ~100 µm. (**Figure 1b**). After the coating was prepared, the structure was analyzed using X-Ray diffraction (XRD). XRD confirmed the semi-crystalline nature of the S-Se coating (**Figure 1c**). The crystalline peaks indexed in XRD corresponds to monoclinic S-Se (blue ink) and hexagonal Se (black ink), respectively whereas the amorphous background signature indicated its amorphous nature.

Next, we investigated tribological properties by friction force microscopy (FFM) to understand the frictional behavior of S-Se/MS and bare MS samples as a function of applied normal load (**Figure 1d** and **Figure S2,** Supplementary Information). Friction was higher overall for the S-Se coating than for bare MS and both experienced a friction increase as the normal load increased. Additionally, both exhibit humidity-independent friction responses suggesting a weak interaction of water with the S-Se surface (**Figure S3,** Supplementary Information). Further, a nano-scratch test was carried out by applying a constant normal load (~3 mN) while measuring the force required to move the tip laterally by 10 $\mu$m. The scratch steadily deepens due to increasing normal load resulting in material pile-up at the end of scratch with negligible side build-up (**Figure 1e**). The maximum lateral force experienced by the indenter is around 1.35 mN (**Figure S4,** Supplementary Information), which is significantly higher than what has been reported for polymer nanocomposite based anticorrosion coatings (*24*). This indicates that the S-Se coating has good wear resistance because it resists scratching, which is reflected in the larger lateral indenter force. We also determined the coefficient of friction (ratio of lateral to normal force) of the S-Se alloy to be ~0.45, which indicates that the S-Se coating exhibits good wear-resistance.

After tribological analysis, we conducted adhesion studies of the S-Se coating onto mild steel surfaces to ensure that delamination would not be a concern under normal operating conditions. Using the pull-off test technique, in which flat counter surfaces were attached onto the coating with epoxy and pulled using a force gauge, we observed cohesive failure with an average cohesion of ~0.37 ± 0.18 MPa (**Figure S5,** Supplementary Information). This demonstrates the coating-steel bond is stronger than the mechanical cohesion within the coating. These results indicate good adhesion properties between the S-Se/MS and that failure would likely be cohesive in nature for the coating.

Next, we conducted a nanoindentation test to understand the elastic and plastic behavior of the S-Se coating. From the load-displacement curve (**Figure 1f**), the hardness and reduced moduli of the coating were calculated to be ~426 MPa and ~8.08 GPa, respectively by Oliver and Pharr method (*25*). It can be inferred from the mapping data that the coating is homogenous over the substrate and the mechanical properties are in line with the single indentation results (**Figure S6, Table S1,** Supplementary Information). The value of the plasticity index (i.e., the ratio of the area enclosed between loading-unloading curves to the loading part of the curve) for S-Se alloy is found to be ~0.6, which is comparable to commercial polymer coatings such as epoxy, polycarbonate, poly (methyl methacrylate), and ultra-high molecular weight polyethylene. While most of these polymers have moduli in the range of 0.5-5 GPa and hardness in the range of 0.1-0.3 GPa in nanoindentation experiments (*21*, *22*), S-Se exhibits markedly higher modulus (~8 GPa) and hardness (~0.43 GPa), while maintaining the plasticity index or ductile nature like viscoelastic polymers (**Figure 1g**). The S-Se alloy developed in this work also exceeds the mechanical properties of polymer nanocomposites which incorporate nano-scale fillers, such as carbon nanotubes[21]. Additionally, while some inorganic coatings such as h-BN and fused silica have higher modulus and hardness, they lack flexibility or ductility like the S-Se alloy. Thus, the unique combinations of tribological and mechanical properties displayed by the S-Se alloy makes it a superior coating when compared to current state-of-the-art anti-corrosive materials.

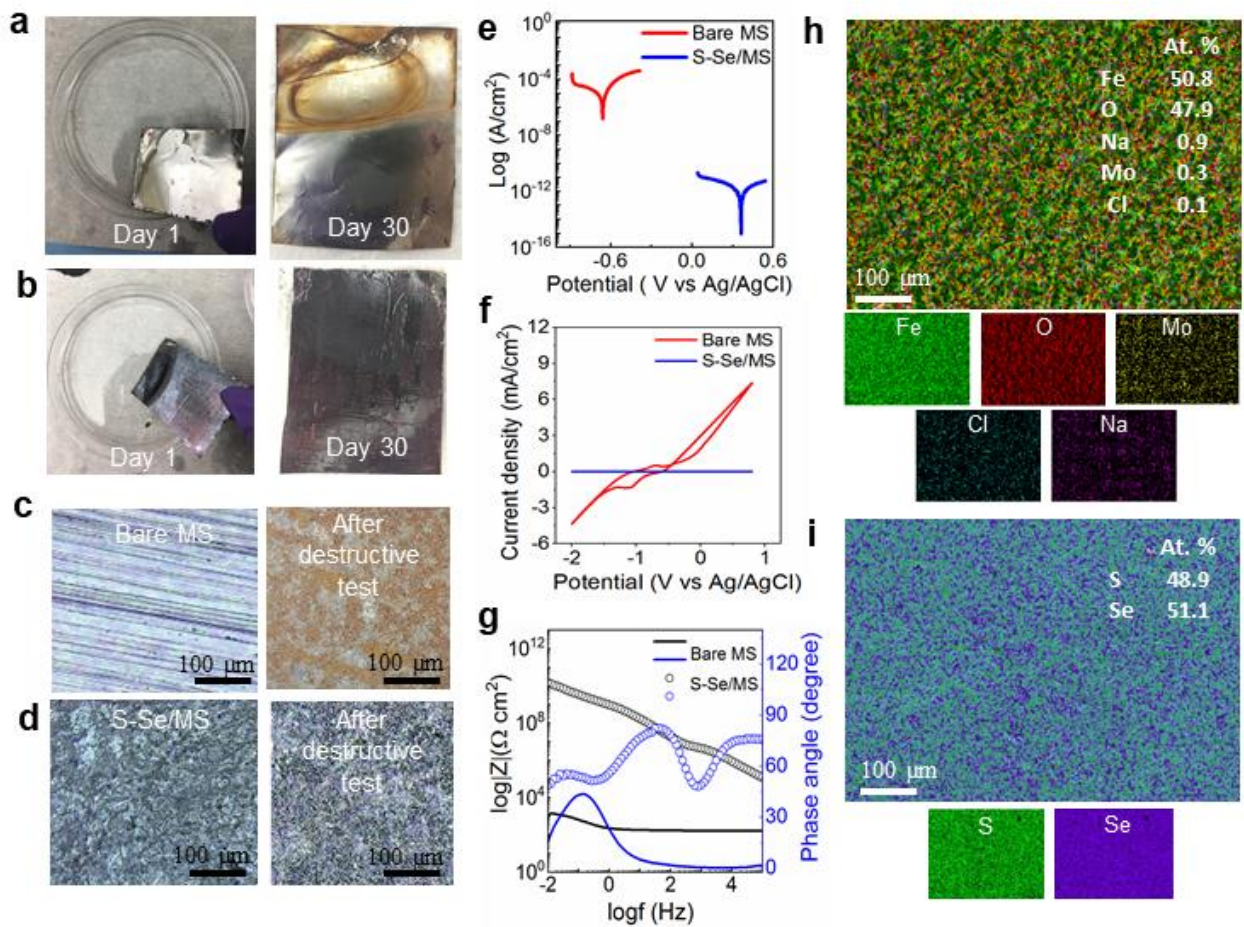

**Figure 2: Corrosion resistance of S-Se coating in simulated seawater (0.6 M NaCl).**

(a) Mild steel (MS) in simulated seawater after day 1 and day 30. The yellow residue in the water is a result of rust formation. Rust is also evident by the formation of brown lines on the MS surface. (b) S-Se coated MS (S-Se/MS) in the seawater on day 1 and day 30. No indication of coating detachment was observed in the residual seawater solution. (c) Pristine bare MS before and after electrochemical destructive test in seawater. (d) S-Se/MS before and after the electrochemical destructive test. The surface of S-Se/MS does not show any signs of deterioration while bare MS shows extensive rust formation. (e-i) Destructive electrochemical test results after ~50 min. of immersion in seawater. (e) Potentiodynamic polarization (Tafel) plots. (f) Cyclic voltammetry measurements. (g) Bode magnitude and phase angle plots. Energy dispersive spectroscopy (EDS) map showing the chemical composition of corrosion products for (h) bare MS and (i) S-Se/MS after exposure to simulated seawater. The S-Se/MS does not exhibit the presence of Fe element indicating barrier protection against electrochemical oxidation.

**Corrosion Resistance in Abiotic Environments**

A promising application of the synthesized S-Se alloy is its use as an oxidation-resistant coating for steel surfaces in marine environments. To study its corrosion protection ability, S-

Se/MS along with bare MS samples were exposed to simulated seawater containing 3.5% or 0.6 M NaCl for ~30 days, which typically is more aggressive compared to natural seawater. The S-Se/MS sample did not exhibit any discoloration or change after 30 days, but the bare MS corroded significantly (visible as yellow stains in the optical image - **Figure 2a**). Further, the S-Se/MS sample did not leave any residual S-Se flakes over a period of 30 days, indicating that the S-Se/MS interface is qualitatively stable and oxidation-resistant (**Figure 2b**). To study the corrosion kinetics of S-Se/MS in simulated seawater, accelerated electrochemical corrosion tests were performed to simulate the effects of aggressive species (chloride and oxygen) that are known to promote cathodic (oxygen reduction) and anodic (Fe oxidation) reactions and accelerate the corrosion of MS. Post-mortem analysis of the degree of corrosion was analyzed using optical imaging. The optical images of bare MS (**Figure 2c**) show a high degree of corrosion attack with a rough (heterogeneous) surface having brown corrosion deposits indicating the formation of rust (i.e., iron oxide-hydroxide), whereas the S-Se/MS (**Figure 2d**) stayed intact with no signs of coating deterioration. We quantitatively analyzed the corrosion resistance of the S-Se/MS using potentiodynamic polarization (PDP) and cyclic voltammetry (CV) techniques. The PDP analysis showed that the S-Se coating suppressed corrosion rates by six orders of magnitude as compared to bare MS (**Figure 2e, Table S2,** Supplementary Information). Further, the PDP or Tafel plots show larger open circuit potential (OCP) value (+363 mV) and significantly lower currents for S-Se/MS compared to bare MS (-657 mV), confirming the passivation behavior of the coatings against corrosive chloride attack. The CV measurements elucidate the impact of the coating as a barrier on chemical reactions involved in a corrosion process. The bare MS showed anodic (-0.70 V vs. Ag/AgCl) and cathodic peaks (-1.71 V vs. Ag/AgCl) with a sharp increase in corrosion current indicating electro-dissolution of iron (Fe) ions (**Figure 2f**). In contrast, S-Se/MS showed a flat line even after 4 CV runs clearly indicating a robust barrier between the electrolytes and MS, thereby preventing corrosion caused by aggressive chloride ions.

Electrochemical Impedance Spectroscopy (EIS) corroborated the strong corrosion barrier properties of the S-Se coating to simulated seawater (**Figure 2g**). In the low frequency (0.01 Hz) region of the Bode impedance plot, S-Se coatings exhibit seven orders of magnitude higher corrosion resistance (~15160 MΩ.cm$^{-2}$) than that of bare MS (~1.2 kΩ.cm$^{-2}$). The phase angle plot for S-Se/MS shows two distinct phase angles peaks, with the first peak representing resistance to penetration of chloride ions through the coating at the electrolyte/coating interface and the second peak attributed to resistance to the electrochemical reaction at the electrolyte/substrate interface. The higher value of phase angle (~90°) at the middle (10 Hz) and high frequency (100 kHz) confirms the absence of electrochemical reaction at the S-Se/MS interface. By contrast, we observed low phase angle maxima (~50°) with a shift to the lower frequency for bare MS. The higher phase angle and impedance values for S-Se/MS denotes a robust coating that lowers the corrosion reaction rates on the surface.

The signatures of corrosion attack by chlorides investigated using EDS analysis reveal the presence of Fe and O peaks on the bare MS surface (**Figure 2h**), suggesting the formation of iron-based corrosion products. The absence of these peaks on the S-Se/MS sample (**Figure 2i**) provides clear evidence that S-Se coatings suppress Fe dissolution. In addition to this, we also conducted electrochemical corrosion analysis in aerated 0.1 M $Na_2SO_4$ salt solution to understand the corrosion resistance properties of S-Se alloy against sulfate salts. The corrosion resistance on S-Se/MS in sulfate medium was six orders of magnitude higher compared to the bare MS (**Figure S7-S8, Table S3,** Supplementary Information). Thus, both the results in chloride and sulfate media demonstrate that S-Se alloy coatings are highly effective in combating corrosion in abiotic chemical environments.

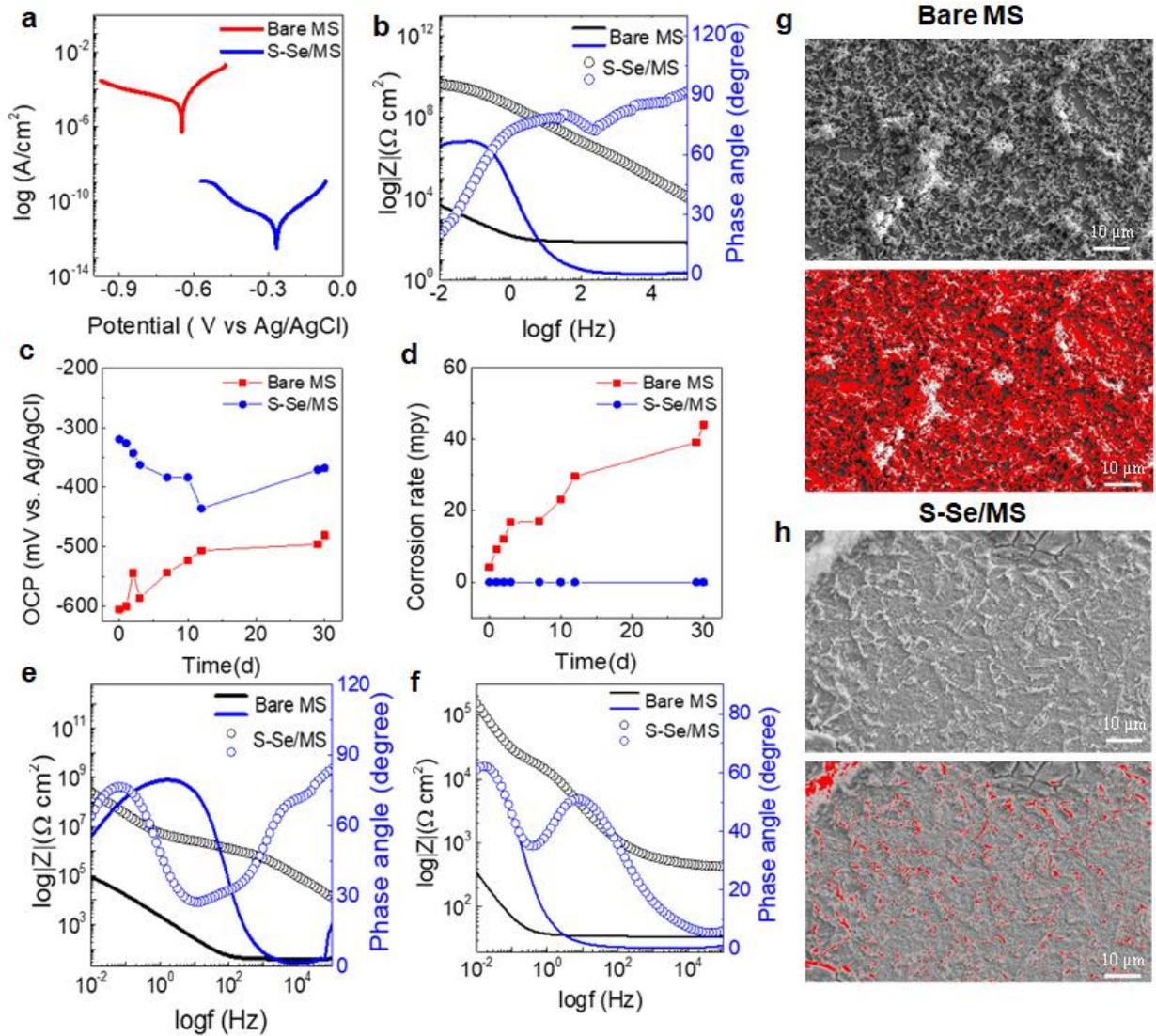

**Figure 3: Corrosion resistance of S-Se coating in biotic environments.**

DC and AC corrosion tests establish microbial corrosion resistance of S-Se/MS exposed to sulfate-reducing bacteria (SRB) '*Desulfovibrio alaskensis* G20' in planktonic condition: (a) Tafel plots and (b) Bode magnitude and phase angle of Bare MS and S-Se/MS after two hours of exposure. DC and AC corrosion tests also establish microbial corrosion resistance of S-Se/MS exposed to SRB in sessile form: Temporal variation of (c) open circuit potential and (d) corrosion rates deduced from linear polarization resistance analysis for 30 days of exposure time. Bode-phase angle plots for bare MS and S-Se/MS for (e) day 1 and (f) day 30. SEM images of (g) bare MS and (h) S-Se/MS. False-colored SEM images (red color denotes rod-shaped *D. alaskensis* cells) are also provided to compare cell attachment.

**Corrosion Resistance in Biotic Environments**

After evaluating the protection of S-Se coatings in abiotic environments, we analyzed the ability of S-Se to suppress aggressive microbially induced corrosion (MIC) involving sulfate-reducing bacteria (SRB) '*Desulfovibrio alaskensis* G20' under anaerobic conditions. SRB is known to accelerate corrosion kinetics by 70 to 90-fold as compared to abiotic corrosion. We evaluated the coating performance in both planktonic (SRB cells are in suspension for 2h) and sessile forms (SRB cells are given 30 days to grow biofilms that are attached to the MS surface).

Tafel analysis and EIS were used to determine the performance of the S-Se coating exposed to planktonic cells within two hours of exposure. The results obtained from Tafel analysis (**Figure 3a**) show that the S-Se coating drastically decreases Fe dissolution due to biogenic $H_2S$ by SRB. The corrosion potential ($E_{corr}$) for S-Se/MS is significantly larger (-280 mV vs -647 mV for bare MS), and its current density ($i_{corr}$) is seven orders of magnitude lower, compared to bare MS (**Table S4,** Supplementary Information). The more positive potential and lower current density of S-Se is a signature of the effective inhibition by the coating against corrosive metabolites ($HS^-$) associated with sulfate reduction. We calculated the inhibiting efficiency ($\eta$) of the S-Se coating to be ~99.99%. The corrosion rate for S-Se/MS (~2.24 x $10^{-4}$ mpy) was five orders of magnitude lower than bare MS (~15.5 mpy). Additionally, Bode magnitude and phase angle plots obtained from the EIS analysis (**Figure 3b**) confirms the barrier property of the coating. The S-Se/MS exposed to planktonic SRB showed six orders of magnitude higher impedance modulus values compared to bare MS in the low-frequency region (~0.01 Hz), while the phase angle for S-Se/MS was almost 90° in the high-frequency region ($10^5$ Hz) in contrast to bare MS. As such, S-Se coatings are highly effective as an anodic barrier to the underlying Fe surface, preventing oxidation kinetics under biotic planktonic conditions.

Compared to planktonic cells, sessile bacteria present in biofilms are known to accelerate corrosion rates by ~100-fold. After establishing the corrosion resistance of S-Se alloy against MIC by the planktonic cells, we further analyzed the corrosion behavior under sessile conditions over 30-day exposure time. Open circuit potential (OCP) for S-Se/MS (**Figure 3c**) is larger (more positive) compared to bare MS throughout the exposure time. For example, the OCP for S-Se/MS on day 1, 12 and 30 was 300, 140 and 112 mV higher compared to bare MS. To quantify the corrosion kinetics on a temporal scale, we used the linear polarization resistance (LPR) method. The corrosion rates obtained from LPR were consistent with the OCP trends. The corrosion rate (maximum value) recorded for S-Se/MS (0.04 mpy on day 30) was 1000 times lower compared to the maximum corrosion rate for bare MS (43.97 mpy on day 30) (**Figure 3d**). As such, the S-Se coating acts as a protective layer and inhibits electrochemical reactions in sessile conditions also.

EIS measurements corroborates the strong corrosion barrier properties of the coatings against SRB biofilms (**Figure 3e).** The low frequency (0.01 Hz) corrosion resistance obtained from Bode plot of S-Se/MS on day 1 is 4 orders of magnitude higher ($10^8$ $\Omega.cm^2$) compared to bare MS ($10^4$ $\Omega.cm^2$). Similarly, on day 30, the corrosion resistance of S-Se/MS was 1000-fold higher compared to bare MS (**Figure 3f**). To observe biofilm growth and assess the degree of MIC resistance, the S-Se/MS and bare MS were removed from the corrosion cell after 30 days of exposure and analyzed using SEM, EDS, and XRD. SEM images of S-Se/MS and bare MS (**Figure 3g, h, Figure S9,** Supplementary Information) clearly indicate that the S-Se coating discourages cell adhesion. The bare MS surface had more densely packed SRB cells, while the S-Se/MS surface contained very few SRB cells in an extra polymeric substance (EPS) matrix. Further, the biofilm matrix was analyzed for the possible formation of corrosion products. The XRD results revealed the presence of different phases of iron oxide, hydroxide and iron sulfide on bare MS (**Figure S10,**

Supplementary Information). In contrast, the S-Se/MS surface did not show any peaks for iron sulfide.

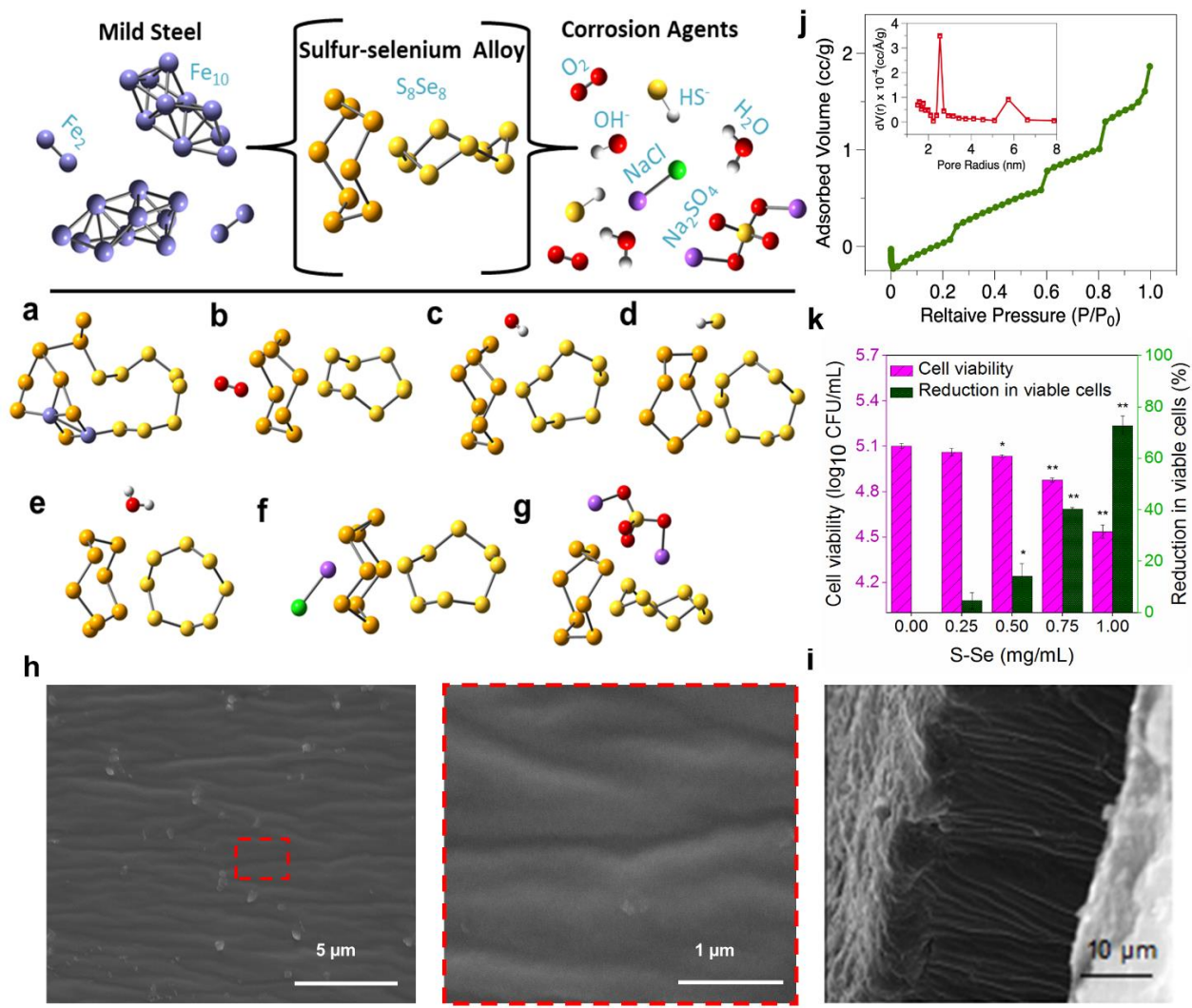

**Figure 4: Corrosion resistance mechanisms of S-Se alloy.**

Iron clusters (Fe$_2$, Fe$_{10}$) at the surface of the reconstructed mild steel (left). Molecular unit for the sulfur-selenium alloy (middle). Corrosion agents O$_2$, OH$^-$, HS$^-$, H$_2$O, NaCl, Na$_2$SO$_4$ (right). Most energetically favorable quantum mechanically optimized geometries resulting from (a) the interaction between Fe$_2$ and S$_8$Se$_8$, (b) the interaction between O$_2$ and S$_8$Se$_8$, (c) the interaction between OH$^-$ and S$_8$Se$_8$, (d) the interaction between HS$^-$ and S$_8$Se$_8$, (e) the interaction between H$_2$O and S$_8$Se$_8$, (f) the interaction between NaCl and S$_8$Se$_8$, (g) the interaction between Na$_2$SO$_4$ and S$_8$Se$_8$. (h) In-plane SEM images of S-Se film coating. The S-Se coating surface is not completely flat as observed in the macroscopic view. (i) Cross-section images of S-Se coating. The coating overall is dense with negligible signs of porosity. (j) Nitrogen adsorption isotherm and pore size distribution of S-Se alloy to quantify the porosity. (k) Antibacterial activity of S-Se alloy against *D. alaskensis* cells assessed using the bacterial plate count method. * P value <0.05 and ** P value <0.01. Bacterial viable count ($\log_{10}$ CFU/mL) and percentage reduction of *D. alaskensis* grown under various concentrations of S-Se alloy.

**Corrosion Resistance Mechanism of the Alloy**

To understand the corrosion mechanism of the S-Se coating, atomic scale DFT simulations were performed to understand the molecular unit interaction between mild steel, S-Se alloy and corrosion agents ($O_2$, $OH^-$, $HS^-$, $H_2O$, NaCl, $Na_2SO_4$). Our previous study has shown, through theoretical and experimental infrared and Raman spectra, that the perpendicular orientation between octa-sulfur and octa-selenium ($S_8Se_8$) depicted in **Figure 4** is the likeliest unit for the S-Se alloy (*23*). There is a very strong dipolar attraction between S and Se, resulting in its insulating nature and high dielectric properties. Further, mild steel used in our experiments contains a very low percentage of carbon atoms which makes it safe to assume that most of the interaction with the S-Se alloy at the interface happens with iron. Our computational model builds on these assumptions by simulating the interaction between the sulfur-selenium molecular unit ($S_8Se_8$) and $Fe_2$ in order to shed light on the chemical nature of the coating properties. We simulated the interaction between the sulfur-selenium unit and corrosive agents such as $O_2$, $OH^-$, $HS^-$, $H_2O$, NaCl, and $Na_2SO_4$ to examine the extent to which a significant degradation of the alloy ensues. We positioned $Fe_2$ and corrosion agents in the proximity of sulfur and selenium atoms to form several configurations and the most energetically favorable configurations were extracted for the computation of Gibbs free energies (**Figure 4a-g**). The difference in Gibbs free energy between products ($Fe_2$-$S_8Se_8$, $O_2$-$S_8Se_8$, $OH^-$-$S_8Se_8$, $HS^-$-$S_8Se_8$, $H_2O$-$S_8Se_8$, NaCl-$S_8Se_8$, $Na_2SO_4$-$S_8Se_8$) and respective reactants taken separately were more favorable for $Fe_2$-$S_8Se_8$ than for $O_2$-$S_8Se_8$, $OH^-$-$S_8Se_8$, $HS^-$-$S_8Se_8$, $H_2O$-$S_8Se_8$, NaCl-$S_8Se_8$ or $Na_2SO_4$-$S_8Se_8$ (-2.422 Hartrees vs -2.040 Hartrees, -2.179 Hartrees, -2.095 Hartrees, -2.035 Hartrees, -2.197 Hartrees, -1.216 Hartrees, respectively). This means that the alloy forms stronger bonds with the reconstructed surface of mild steel than it does with corrosion agents. Except for sodium sulfate ($Na_2SO_4$), water ($H_2O$) appears to be the least corrosive of the corrosion agents because it has the second highest free energy difference between the products and reactants (-2.035 Hartrees). Besides NaCl, bisulfide (HS) appears to be the most corrosive of the corrosion agents because it has the lowest free energy difference between the products and reactants (-2.179 Hartrees) in accordance with its well-known highly toxic and corrosive properties. It is also worth noting that the geometry of the molecular unit for S–Se is highly disrupted when interacting with $Fe_2$ (**Figure 4a**). Indeed, $Fe_2$ breaks both the S ring and the Se ring to form a highly disordered $Fe_2$-$S_8Se_8$ cluster in agreement with the experimentally observed amorphous layer at the interface between the alloy and the surface of mild steel. In a diametrically opposite manner, the geometry of the molecular unit for the alloy is not disrupted when interacting with corrosion agents $O_2$, $OH^-$, $HS^-$, $H_2O$, NaCl, $Na_2SO_4$ (**Figure 4b-g**). Two observations are worth underscoring here. First, the ring-like structure of the alloy unit as well as the perpendicular orientation of the S ring and the Se ring are not significantly altered as a result of the interaction with the corrosive agents. Second, no intercalation of the corrosive agents into the S ring or the Se ring is observed. This is especially true for sodium sulfate that exhibits the highest free energy difference (-1.216 Hartrees). These two observations shed light on the chemical nature of the protective anticorrosive properties of the S-Se alloy and indicate that it is highly unfavorable for corrosion agents to permeate through the alloy units. Taken together, our simulations illustrate, from a molecular perspective, two distinct behaviors of the alloy. On one hand, the alloy shows strong adhesion to the surface of mild steel thus demonstrating its coating capabilities. On the other hand, the alloy exhibits great resistance to corrosive agents thus demonstrating its anticorrosive properties.

Beside DFT simulations, we have experimentally analyzed the morphology of the S-Se coating to understand the corrosion protection mechanism. We analyzed the morphology of S-Se alloy using SEM to determine the porosity in the surface, since porosity will have an adverse effect on the ability of the film to block the penetration of corrosive species. The in-plane SEM images of S-Se alloy show a micro-scale uneven surface, although it appears completely flat from the macroscopic scale (**Figure 4h**). However, even at the highest magnification, there is no indication of any porosity present in the structure; thus, validating the previous ab initio calculations. For cross-section images too, we did not observe any signs of porosity (**Figure 4i**). Besides microscopic observations, $N_2$ adsorption isotherms of the S-Se alloy were investigated and the multi-point BET surface area was calculated (**Figure 4j**). The S-Se alloy has a very low BET surface area (~0.25 $m^2/g$) that further confirms the non-porous microstructure of the alloy. Moreover, from the pore size distribution and pore volume analyses, we found that the alloy exhibits negligible pore volume, available for gas adsorption. Furthermore, XRD revealed our coating as a semi-crystalline material that lacks structural defects such as grain boundaries or dislocations (**Figure 1c**). This facilitates the formation of a relatively defect-free interface with the corrosive environment.

Finally, we performed antibacterial activity studies to understand the nature of the interaction of S-Se alloy in biotic environments. Both S and Se have anti-bacterial and anti-microbial properties. Hence, the S-Se alloy is also likely to have antibacterial properties. Antibacterial activity of Sulfur-selenium alloy against *D. alaskensis* was assessed using bacterial plate count method on lactate C medium agar plates. Results showed that the percent reduction of *D. alaskensis* cells increased with increasing concentrations of S-Se (**Figure 4k**). *D. alaskensis* cells treated with 1 mg/ml showed maximum percent reduction of 70% followed by 0.75 mg/ml that showed a 39.8% percent reduction, which confirms that the S-Se alloy exhibits significant antibacterial activity. This explains its potency (see **Figure 3**) in suppressing biofilm formation and preventing microbial corrosion in biotic environments.

**Conclusions**

In summary, unwanted metal corrosion costs industries across the world billions of dollars annually, and researchers have sought materials and methods to combat this issue for decades. However, many current advancements in anti-corrosive coatings include prohibitively expensive materials or non-scalable processing methods that are not versatile enough to prevent corrosion in many different environments. In this study, we demonstrate S-Se alloy as a universal anti-corrosive coating, which is easily processed, mechanically robust, and effective in both biotic and abiotic environments. The S-Se coating displays moduli and hardness values far greater than most polymer and nanocomposite coatings, while also maintaining its ductile and elastic recovery nature unlike inorganic coatings. S-Se alloy coatings are able to protect mild steel from diverse corrosive environments due to their unique combination of properties, which includes their insulating and impermeable nature, high coating capacitance, and intrinsic antimicrobial properties. We believe this study represents an important contribution to current high-performance coatings research as a low-cost "all-in-one" anti-corrosion coating.

**Acknowledgements:** The authors would like to acknowledge National Science Foundation CAREER award (#1454102), NSF RII T-1 FEC award #1849206, NSF RII T-2 FEC award# 1920954, and NASA (# NNX16AQ98A) for funding this research. **Author contributions:** M.M.R. and P.M.A. designed and coordinated the study, S.S. and Y.C. synthesized and characterized S-Se alloy, G.C., S.C., J.R.K. and G.V. performed corrosion studies, T.A., A.H., A.B.P., P.M.S. and T.F. performed and analyzed nanomechanical testing, T.T. performed computational studies, M.B. carried out rheology testing, M.M.R., P.M.A, G.C., G.V., R.V., T.A., N.K. cowrote the paper. M.M.R., P.M.A., N.K., T.F. and G.V. discussed the results and commented on the manuscript. **Competing Financial Interests**: The authors declare no competing financial interests. **Data and materials availability**: Methods and other Supplementary Information is available in the online version of the paper. Reprints and permissions information is available online. Correspondence and requests for materials should be addressed to M.M.R.